\documentclass[submission,copyright,creativecommons]{eptcs}
\providecommand{\event}{QAPL 2017} 
\usepackage{underscore}           
\usepackage{graphicx}
\usepackage{amsmath}
\usepackage{amsfonts}
\usepackage{amsthm}
\usepackage{tikz}
\usepackage{pgfplots}
\usepackage{multirow}
\usepackage{xspace}
\usepackage{color, colortbl}
\usepackage{rwthcolors}
\usepackage{subcaption}
\usepackage[linesnumbered]{algorithm2e}
\usepackage{hyperref}
\usepackage{listings}
\usepackage[siunitx]{circuitikz}
\usetikzlibrary{arrows,shadows,petri,automata,patterns}
\usetikzlibrary{decorations.pathmorphing}
\usetikzlibrary{decorations.markings}
\usetikzlibrary{positioning,shapes,calc, decorations.pathreplacing, arrows, backgrounds, fit, shapes.symbols}
\usepackage{booktabs}

\usetikzlibrary{calc}
\usetikzlibrary{intersections}
\usepackage{todonotes}
\usepackage{wrapfig}

\newcommand{\Loc}{\ensuremath{\mathit{Loc}}\xspace}
\newcommand{\Var}{\ensuremath{X}\xspace}

\newcommand{\Edge}{\ensuremath{\mathit{Edge}}\xspace}

\newcommand{\Inv}{\ensuremath{\mathit{Inv}}\xspace}
\newcommand{\Init}{\ensuremath{\mathit{Init}}\xspace}
\newcommand{\Flow}{\mathit{Flow}}
\newcommand{\Pred}[1]{\mathit{Pred}_{#1}}

\renewcommand{\H}{\ensuremath{\mathcal{H}}}
\newcommand{\hypro}{\textsc{HyPro}\xspace}
\newcommand{\spaceex}{\textsc{SpaceEx}\xspace}
\newcommand{\flowstar}{\textsc{Flow}$^*$\xspace}
\newcommand{\glpk}{\textsc{GLPK}\xspace}
\newcommand{\smtrat}{\textsc{SMTRAT}\xspace}
\newcommand{\soplex}{\textsc{SoPlex}\xspace}
\newcommand{\zthree}{\textsc{Z3}\xspace}
\newcommand{\cora}{\textsc{Cora}\xspace}
\newcommand{\hycreate}{\textsc{HyCreate}\xspace}
\newcommand{\hyreach}{\textsc{HyReach}\xspace}
\newcommand{\soapbox}{\textsc{SoapBox}\xspace}
\newcommand{\ppl}{\textsc{PPL}\xspace}
\newcommand{\var}{\ensuremath{V}}
\newcommand{\vact}{\ensuremath{\var_{\textit{act}}}}
\newcommand{\vsen}{\ensuremath{\var_{\textit{sen}}}}
\newcommand{\vdyn}{\ensuremath{\var_{\textit{cont}}}}
\newcommand{\vin}{\ensuremath{\var_{\textit{in}}}}
\newcommand{\vout}{\ensuremath{\var_{\textit{out}}}}
\newcommand{\vloc}{\ensuremath{\var_{\textit{loc}}}}
\newcommand{\temperature}{\ensuremath{T}}

\newcommand{\disc}{\textit{disc}}
\newcommand{\clock}{\textit{clock}}
\newcommand{\rest}{\textit{rest}}
\newcommand{\dVar}{\Var_{\disc}}
\newcommand{\cVar}{\Var_{\clock}}
\newcommand{\rVar}{\Var_{\rest}}
\newcommand{\isFirst}{\textit{isFirst}}

\title{Divide and Conquer: Variable Set Separation in Hybrid Systems Reachability Analysis\footnote{This work was partially supported by
the German Research Council (DFG) in the context of the HyPro
project.}
}
\author{Stefan Schupp \qquad Johanna Nellen \qquad Erika \'Abrah\'am
\institute{RWTH Aachen University, Germany}
\email{\{ stefan.schupp | johanna.nellen | abraham \}@cs.rwth-aachen.de}}
\def\titlerunning{Variable Separation for Hybrid Systems Reachability}
\def\authorrunning{S.Schupp, J.Nellen \& E.\'Abrah\'am}
\begin{document}
\maketitle

\newtheorem{definition}{Definition}

\begin{abstract}
  In this paper we propose an improvement for
  flowpipe-construction-based reachability analysis techniques for
  hybrid systems. Such methods apply iterative successor computations
  to pave the reachable region of the state space by state sets in an
  over-approximative manner. As the computational costs steeply increase with the dimension, in this work we
  analyse the possibilities for improving scalability by dividing the
  search space in sub-spaces and execute reachability computations in
  the sub-spaces instead of the global space. We formalise such an
  algorithm and provide experimental evaluations to compare the
  efficiency as well as the precision of our sub-space search to the original search in the global space.
\end{abstract}

\section{Introduction} \label{sec:introduction}

The increasing usage of digital control for safety-critical dynamical
systems has resulted in an increasing need for formal verification
approaches for \emph{hybrid systems}, i.e., for systems with mixed
discrete-continuous behaviour, which are often modelled as
\emph{hybrid automata}. Due to intensive research, nowadays several
approaches and tools exist for the \emph{reachability analysis} of
hybrid automata.  As the reachability problem for hybrid automata is
in general undecidable, most approaches compute an
\emph{over-approxima\-tion} of the set of states that are reachable in a
given hybrid automaton model. Due to the over-approxima\-tion, these
techniques can be used to prove the safety of system models, i.e., the
fact that a given set of unsafe states is not reachable in the model,
but they cannot be used to prove unsafety.

In this work we focus on \emph{flowpipe-construction-based}
reachability analysis techniques. These techniques use certain data
types to \emph{represent} state sets, whereas each representation has
its strengths and weaknesses in terms of precision, memory
requirements, and efficiency of certain operations on them which are
needed for the reachability computations. The reachability analysis starts from an initial
state set and iteratively over-approximates successors by further state
sets.  For a given set of states, the successors via a discrete computation
step (\emph{jump}) are over-approximated by a single set, the successors via
time evolution (the so-called \emph{flowpipe}) are covered in an
over-approximative manner by a sequence of state sets.

Unfortunately, these successor computations often lead to either
strong over-approximations or high computational costs. Though
state-of-the-art tools like \spaceex \cite{spaceex}, \flowstar
\cite{flowstar}, or \hypro \cite{hypro} can already successfully verify
a wide range of challenging applications, they still have problems to
analyse large models with complex behaviour. Such models arise for
example from applications, where a physical or chemical plant is
controlled by a discrete controller. Our focus is on digital control
by programs running on \emph{programmable logic controllers}
(\emph{PLCs}). To build a formal model of such a system, the PLCs, the
programs running on them, the dynamic plant behaviour, and the interactions
between these components can be modelled by a hybrid automaton,
to which available reachability analysis
tools can be applied. For practically relevant systems, however, the
size of the resulting composed models often exceeds the capabilities of
state-of-the-art tools.

Whereas general techniques to increase the scalability of reachability
analysis are hard to develop, for dedicated model types there might be
some hot spots. Models of PLC-controlled plants have some specific
properties we can exploit: Firstly, they possess a relevant number of
\emph{discrete variables}. Secondly, some actions are triggered by
deadlines, modelled by the values of \emph{clocks} along with
corresponding thresholds. Thirdly, the evolution of some physical
quantities might depend on the time \emph{linearly}, others not. In
standard reachability analysis, these model parts are handled
uniquely. In this paper we propose to split the state space into
several sub-spaces, between which the dependence is loose enough to
execute successor computations independently. Though this procedure
leads to additional over-approximation, the error can be reduced.
Furthermore, we show on some experiments that this additional
over-approximation is often minor and is well compensated by the
reduced computational requirements.

We are aware of the work \cite{chen:decomposed} that is closely related to the work described in this paper. The authors of \cite{chen:decomposed} also use variable set separation and computations in sub-spaces, but with two main differences. On the one hand, the work \cite{chen:decomposed} is more general as they allow also closer dependencies between the sub-spaces than we can support. On the other hand, their work is restricted to Taylor models, whereas our approach is applicable to any state set representation type.

\emph{Overview}\quad After providing some preliminaries in Section
\ref{sec:preliminaries} and a description of our \hypro programming
library in Section \ref{sec:hypro}, in Section \ref{sec:setSeparation}
we describe our method for the separation of variable dimensions and
the modified reachability algorithm. We provide some experimental
results in Section \ref{sec:experiments} before we conclude the paper
in Section \ref{sec:conclusion}.

\section{Preliminaries}
\label{sec:preliminaries}


\noindent\textbf{Hybrid automata}\quad For a given set $X=\{x_1,\ldots,x_d\}$ of variables let $\Pred{X}$ be
the set of all quantifier-free arithmetic predicates with free
variables from $X$, using the standard syntax and semantics over the
real domain.  We use the notation
$\dot{X}=\{\dot{x}_1,\ldots,\dot{x}_d\}$ to represent first
derivatives and $\Var'=\{x_1',\ldots,x_d'\}$ to represent the result
of discrete resets of variable values.  Sometimes we also see the
variable space as the $d$-dimensional real space and use the vector
notation $x=(x_1,\ldots,x_d)\in\mathbb{R}^d$.

\begin{definition}[\cite{henzinger:hybrid}]
	A \emph{hybrid automaton} 
	$\H = (\Loc,\Var,\Flow,\Inv,\Edge,\Init)$ is a tuple specifying
	\begin{itemize}
		\item a finite set $\Loc$ of \emph{locations} or \emph{control modes};
		\item a finite ordered set $\Var=\{x_1,\ldots,x_d\}$
                  of real-valued \emph{variables}, where
                  $d$ is the \emph{dimension} of
                  $\H$;
		\item for each location its \emph{flow} or \emph{dynamics} by the function $\Flow:\Loc\rightarrow\Pred{\Var\cup\dot{\Var}}$;
		\item for each location an \emph{invariant} by the function $\Inv:\Loc\rightarrow\Pred{\Var}$;
		\item a finite set $\Edge \subseteq \Loc\times
                  \Pred{\Var}\times\Pred{\Var\cup\Var'}\times\Loc$ of \emph{discrete transitions} or
                  \emph{jumps}. For a jump
                  $(l_1,g,r,l_2)\in\Edge$, $l_1$ is its
                  \emph{source} location, $l_2$ is its \emph{target}
                  location, $g$ specifies the jump's \emph{guard}, and
                  $r$ its \emph{reset} function;
		\item an \emph{initial} predicate for each location by the function $\Init:\Loc\rightarrow\Pred{\Var}$.
	\end{itemize}
	\label{def:hybridAutomaton}
\end{definition}

In this paper we consider only \emph{autonomous linear} hybrid
automata whose initial conditions, invariants and jump guards are
linear and can be written in the form $Ax\leq b$ (where
$x=(x_1,\ldots,x_d)$ are the model variables, $A$ is a $d\times d$
matrix and $b$ a $d$-dimensional vector), whose jump resets are also
linear and can be written as $x'=Ax$, and whose flows are defined by conjunctions
of linear \emph{ordinary differential equations} (\emph{ODEs}), which
can be written\footnote{Note that
  ODEs of the form $\dot{x}=Ax+b$ can be also encoded without a $b$
  component on the cost of new variables with zero derivatives. A similar approach is possible for jump resets of the form $x'=Ax+b$.} as 
$\dot{x}=Ax$.
Note that such automata allow only linear predicates, whose solutions
are convex polytopes.

A \emph{state} $(l,x)\in\Loc\times\mathbb{R}^d$ of a hybrid automaton
specifies the location $l\in\Loc$ in which the control resides and the
current values $x\in\mathbb{R}^d$ of the variables. For
$p\subseteq\mathbb{R}^d$, by $(l,p)$ we denote the state set $\{(l,x)\
| \ x\in p\}$. An \emph{execution}
$(l_0,x_0)\stackrel{t_0}{\rightarrow}(l_1,x_1)\stackrel{e_1}{\rightarrow}(l_2,x_2)\stackrel{t_2}{\rightarrow}\ldots$
of a hybrid automaton starts in an initial state $(l_0,x_0)$ such that
$x_0$ satisfies $\Init(l_0)$, and executes a sequence of alternating
continuous and discrete steps. A continuous step
$(l_i,x_i)\stackrel{t_i}{\rightarrow}(l_{i+1},x_{i+1})$ with
$l_i=l_{i+1}$ models time evolution: starting from $x_i$, the variable
values evolve according to the flow (ODEs) $\Flow(l_i)$ of the current
location for $t_i$ time units, where the location's invariant must hold
during the whole duration of the step. A discrete step
$(l_i,x_i)\stackrel{e_i}{\rightarrow}(l_{i+1},x_{i+1})$ typically
models controller execution: if the source of a jump $e_i$ is $l_i$,
the guard of $e_i$ is satisfied by $x_i$, the reset predicate of $e_i$
is satisfied by $(x_i,x_{i+1})$, and $l_{i+1}$'s invariant is true for
$x_{i+1}$ then the jump $e_i$ can be taken, moving the control from
$l_i$ to $l_{i+1}$ with resulting variable values $x_{i+1}$.

A state of a hybrid automaton $\H$ is called \emph{reachable} if there is an execution leading to it. Given a set $T$ of unsafe states, $\H$ is called \emph{safe} if no state from $T$ is reachable in $\H$.

Hybrid automata can be composed using \emph{parallel composition},
which we do not define here formally. Intuitively, jumps in different
components can be synchronised (using synchronisation labels) if they
should take place simultaneously, whereas local computation steps can
also be executed in isolation; time evolves simultaneously in all components.

\smallskip
\noindent\textbf{Reachability analysis}\quad
The \emph{reachability problem} for hybrid automata is the problem to
decide whether a given state (or any state from a given set) is
reachable in a hybrid automaton.  As the reachability problem for
hybrid automata is in general undecidable, some approaches aim at
computing an \emph{over-approximation} of the set of reachable states
of a given hybrid automaton. We focus on approaches based on
\emph{flowpipe construction}, which iteratively over-approximate the
set of reachable states by the union of a set of state sets. To
\emph{represent} a state set, typically either a \emph{geometric} or a
\emph{symbolic} representation is used. Geometric representations
specify state sets by geometric objects like boxes, (convex)
polytopes, zonotopes, or ellipsoids, whereas symbolic representations
use, e.g., support functions or Taylor models. These representations
might have major differences in the precision of the representation
(the size of over-approximation), the memory requirements and the
computational effort needed to apply operations like intersection,
union, linear transformation, Minkowski sum or test for emptiness.
For example, boxes 
perform well in terms of computational effort for set
operations, but usually introduce a large over-approximation error.
Thus the choice of the representation is a compromise between
the advantages and disadvantages regarding these measures.

Before the reachability analysis starts, all predicates
$\varphi\in\Pred{X}$ in the respective linear hybrid automaton
(initial predicates, invariants, jump guards) as well as the unsafe
state set need to be represented in some state set representation
(usually the same representation for all predicates). Furthermore, the
jump resets need to be formalised as linear transformations.

For a given state set $p$, flowpipe-construction-based approaches
compute the successors of the states from $p$ by first
over-approximating the set of states reachable via time evolution
(\emph{flowpipe}) and afterwards the set of states reachable from the
flowpipe as jump successors. Time evolution is usually restricted to a
\emph{time horizon} (either per location or for the whole execution),
which is divided into smaller time steps. The states reachable from
$p$ via one time step are over-approximated by a state set $p_1$,
for which again the time successors $p_2$ via one time step are
computed. This procedure is repeated until the time horizon is reached
or the successor set gets empty (due to the violation of the current
location's invariant). The union of the resulting state sets
$p_1,\ldots,p_k$, which are called \emph{flowpipe segments},
over-approximates the flowpipe. For each outgoing jump and each of the
flowpipe segments the jump successors are computed, to which the above
procedure is applied iteratively until a given upper bound (\emph{jump
  depth}) on the number of jumps is reached or until a fixed point is
detected. Thus the reachability computation results in a \emph{search
  tree} with state sets as nodes. In order to reduce the computational
effort, \emph{clustering} and \emph{aggregation} can be applied to
over-approximate the successors of the flowpipe segments for a given
jump by a fewer number of segments respectively by a single state set.

\smallskip
\noindent\textbf{Programmable logic controllers}\quad
\emph{Programmable logic controllers} (\emph{PLCs}) are digital
controllers widely used in industrial applications, for instance in
production chains. A PLC has input and output pins that are connected
with the sensors and the actuators of a plant. Control programs
running on a PLC specify the output of the PLC in dependence of its
input. These control programs are executed in a cyclic manner.
First, the PLC \emph{reads} the current state of the sensors and the
actuators of the plant and stores this information in input
  registers.  Next all programs on the PLC \emph{execute} in parallel
to compute the next output values based on the last input, and store
the results in some output registers.  These computations might
use local variables, stored in some local registers.  In the
last step of the cycle the PLC \emph{writes} the computed output
values to the output pins that are connected to the actuators of the
plant. In contrast to some implementations that assure a cycle
duration within a time interval, for simplicity in this work we assume
a constant cycle time (however, our approach can be easily extended to
interval durations).

\begin{figure}[t]
	\begin{center}
	 	\begin{minipage}[c]{0.7\textwidth}
	 	 	\scalebox{0.7}{\begin{tikzpicture}[font=\normalsize, >=latex]

  \node[draw, rectangle, minimum width=2.5cm] (vact) at(2.5,0.75){\begin{tabular}{c}Actuators\\$\vact$\end{tabular}};
  \node[draw, rectangle, minimum width=2.5cm] (vsen) at(2.5,-0.75){\begin{tabular}{c}Sensors\\$\vsen$\end{tabular}};
  \node[draw, rectangle, minimum width=2.5cm] (vdyn) at(-1,0){\begin{tabular}{c} Physical\\quantities\\$\vdyn$\end{tabular}};
  \node[draw=none,fill=none] (plant) at (-1.8,1.5) {\textbf{Plant}};
  \draw[<->] ($(vdyn.east)+(0,0.1)$) --++ (0.5,0) |- (vact.west);
  \draw[<->] ($(vdyn.east)+(0,-0.1)$) -++ (0.5,0) |- (vsen.west);

\begin{scope}[xshift=1cm]
  \node[] (plc) at (5.5,1.5){\textbf{PLC}};
  \node[] (sfc) at (10,1.25){\textbf{Programs}};
  \node[draw, rectangle, minimum width=2cm] (vin) at(7.5,0.75){\begin{tabular}{c}Input\\$\vin$\end{tabular}};
  \node[draw, rectangle, minimum width=2cm] (vout) at(7.5,-0.75){\begin{tabular}{c}Output\\$\vout$\end{tabular}};
  \node[draw, rectangle] (vloc) at(10,0){\begin{tabular}{c} Computation\\$\vloc$\end{tabular}};
  \coordinate (read) at ($(vin.west)+(-1.5,0.0)$) {};
  \coordinate (write) at ($(vout.west)+(-1.5,0.0)$) {};
  \draw[->] (read) -- node[midway, above]{\small\textsf{read}} ($(vin.west)+(-0.1,0.0)$);
  \draw[->] ($(vout.west)+(-0.1,0)$) -- node[midway, above]{\small\textsf{write}} (write);  

  \draw[->, thick] ($(vact.east)+(0,0.2)$) -- ($(read)+(-0.1,0.2)$);
  \draw[->, thick] (vsen.east) --++ (0.3,0) -- ($(read)+((-0.4,-0.2)$) -- ($(read)+((-0.1,-0.2)$);
  \draw[<-, thick] ($(vact.east)+(0,-0.2)$) --++ (0.3,0) -- ($(write)+((-0.4,0)$) -- ($(write)+(-0.1,0)$);
\end{scope}
  
  \begin{pgfonlayer}{background}
	\filldraw[green25, draw=black100] ($(plant.north)+(0,0.2)$) -| ($(vact.east)+(0.2,0)$) |- ($(vsen.south)+(0,-0.3)$) -| ($(vdyn.west)+(-0.2,0)$) |- ($(plant.north)+(0,0.2)$);  
	\filldraw[blue25, draw=black100] ($(plc.north)+(0,0.2)$) -| ($(vloc.east)+(0.3,0)$) |- ($(vout.south)+(0,-0.3)$) -| ($(plc.west)+(-0.2,0)$) |- ($(plc.north)+(0,0.2)$);
	\filldraw[blue25, draw=black100] ($(sfc.north)+(0,0.0)$) -| ($(vloc.east)+(0.1,0)$) |- ($(vout.south)+(0,-0.1)$) -| ($(vout.west)+(-0.1,0)$) |- ($(sfc.north)+(0,0.0)$);  
  \end{pgfonlayer}
  
\end{tikzpicture}%
}
	  	\end{minipage}
	\hfill
	 	\begin{minipage}[c]{0.28\textwidth}
	  	\scalebox{0.45}{\tikzstyle{loc}=[circle, text ragged, minimum width = 1.5cm, rounded corners, draw, thick, inner sep=2pt]
\tikzstyle{io} = [rectangle, fill, minimum width=0.25cm, minimum height=0.5cm]
 
\begin{tikzpicture}[node distance = 5.2cm and 1.2cm, font=\large, >=latex]
\begin{scope}
  \node [loc] (c0) at (2.5,2) {
  		\begin{tikzpicture}
  			\draw[->](0.0,0.0) -- (1.2,0.0);
 			\draw[->](0.0,0.0) -- (0.0,0.8);
 			\draw[-] (0.0,0.2) -- (0.2,0.2);
 			\draw[-] (0.2,0.2) -- (0.2,0.6);
 			\draw[-] (0.2,0.6) -- (0.4,0.6);
 			\draw[-] (0.4,0.6) -- (0.4,0.2);
 			\draw[-] (0.4,0.2) -- (0.6,0.2);
 			\draw[-] (0.6,0.2) -- (0.6,0.6);
 			\draw[-] (0.6,0.6) -- (0.8,0.6);
 			\draw[-] (0.8,0.6) -- (0.8,0.2);
 			\draw[-] (0.8,0.2) -- (1.0,0.2);			
  		\end{tikzpicture}
 	};
  \node [loc] (c1) at (5.5,2) {
  	\begin{tikzpicture}
		\draw[->](0.0,0.0) -- (1.2,0.0);
		\draw[->](0.0,0.0) -- (0.0,0.8);
		\draw[-] (0.0,0.6) -- (0.4,0.6);
		\draw[-] (0.4,0.6) -- (0.4,0.2);
		\draw[-] (0.4,0.2) -- (0.6,0.2);
		\draw[-] (0.6,0.2) -- (0.6,0.6);
		\draw[-] (0.6,0.6) -- (0.8,0.6);
		\draw[-] (0.8,0.6) -- (0.8,0.2);
		\draw[-] (0.8,0.2) -- (1.0,0.2);			
	\end{tikzpicture}
  };

  \node[io] (controllerOut) at(7.5,2) {};
  \node[io] (controllerIn)  at(0.5,2) {};

  \draw[->,thick, bend left] (c0) to (c1);
  \draw[->,thick, bend right] (c0) to (c1);
 
  \draw[->,shorten <=0.1cm] (controllerIn) to (c0);
  \draw[->,shorten >=0.1cm] (c1) to (controllerOut);

    \node[loc] (l1) at (4,-2){
	  	\begin{tikzpicture}
			\draw[->](0.0,0.0) -- (1.2,0.0);
			\draw[->](0.0,0.0) -- (0.0,0.8);
			\draw (0,0) parabola bend (0,0) (1,0.6);
		\end{tikzpicture}
   	};
    \node[loc] (l2) at (4,-4) {
      	\begin{tikzpicture}
		    \draw[->](0.0,0.0) -- (1.2,0.0);
		    \draw[->](0.0,0.0) -- (0.0,0.8);
			\draw (0,0.6) parabola bend (0,0.6) (1,0.2);
		\end{tikzpicture}
	};

    \node[io] (plantOut) at(1.5,-3) {};
    \node[io] (plantIn)  at(6.5,-3) {};

    \draw[->, very thick] (controllerOut) --++(1,0)|- (plantIn);
    
    \draw[->,shorten <=0.1cm] (plantIn.west) to (l1.east);
    \draw[->,shorten <=0.1cm] (plantIn.west) to (l2.east);
    \draw[->,shorten >=0.1cm] (l1.west) to (plantOut);
    \draw[->,shorten >=0.1cm] (l2.west) to (plantOut);
    
    \draw[->, very thick] (plantOut) -| ($(controllerIn.west)+(-1,0)$) -- (controllerIn.west);
 
    \draw[->, bend left] (l1) to (l2);
    \draw[->, bend left] (l2) to (l1);
 
\node[] (controller) at ($(c0.north) + (1.5,0.5)$) {\textbf{Controller}};
 
\node[] (plant) at ($(l1.north) + (0,.5)$) {\textbf{Plant}};
 
\begin{pgfonlayer}{background}
  \filldraw[blue25, draw=black100] (controllerIn.west) |- ($(controller.north)+(0,0.25cm)$) -| (controllerOut.east) |- ($(c0.south)+(0,-0.5cm)$) -| (controllerIn.west);
  \filldraw[green25, draw=black100] (plantOut.west) |- ($(plant.north)+(0,0.25cm)$) -| (plantIn.east) |- 
($(l2.south)+(0,-0.25cm)$) -| (plantOut.west);
\end{pgfonlayer}
     
\end{scope}
 
\end{tikzpicture}}
	  	\end{minipage}
	\end{center}
	\caption{PLC controller: Interface between plant and controller (left) and cyclic execution model (right).}
	\label{fig:plantInterface}
\end{figure}
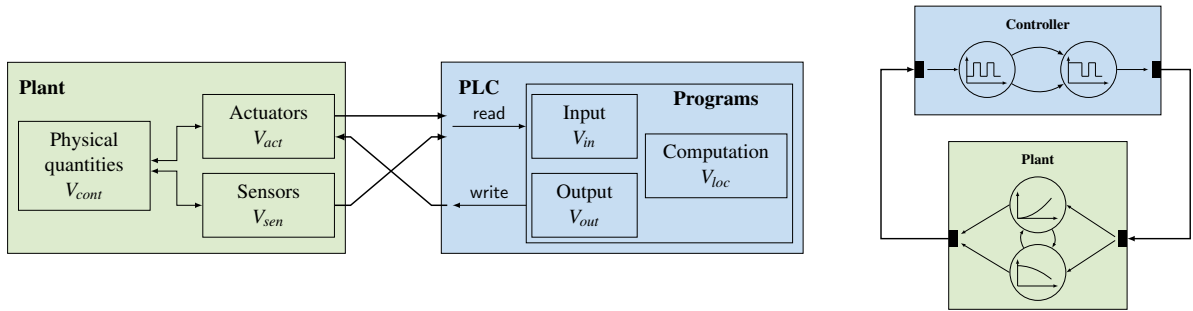

To model a plant we introduce variable sets $\vdyn, \vact,$ and
$\vsen$ to represent the state of physical quantities, the actuators,
respectively the sensors (see left of Figure
\ref{fig:plantInterface}). For the modelling of a controller we use
sets $\vin$, $\vout$, and $\vloc$ of variables to represent the PLC
registers for input, output respectively local variables.
Additionally, we need one variable (clock) per PLC to account for the
PLC cycle time.

A schematic overview of the hybrid automaton we use to model
PLC-controlled plants is shown in the right of Figure
\ref{fig:plantInterface}. We could model the system by specifying
hybrid automata models for the plant, the PLC, and the programs running
on the PLC, and compose them using label synchronisation to model
synchronous events in the PLC cycle. However, these models allow heavy
interleaving between continuous time evolution and discrete PLC
computation steps, leading to models that pose a challenge for
reachability analysis tools. Therefore, we make use of the fact that
the PLC execution between reading the input and writing the output has
no influence on the plant's state: we model the plant evolution and
the concurrent cyclic PLC execution by toggling between a controller
model and a plant model, assuming that all controller actions are
executed instantaneously after the input is read, the plant evolves
for the duration of the PLC cycle, and the output is written at the
end of the cycle.  We refer to \cite{phdNellen} for more information
on the modelling of PLC-controlled plants.

\section{The HyPro Library}\label{sec:hypro}

\begin{figure}[t]
	\centering
\scalebox{0.8}{
	\begin{tikzpicture}[>=latex, class/.style={draw,rectangle,thick, minimum height=.5cm, text width=3.5cm, anchor=west, fill=white}, implements/.style={-open triangle 45, dashed},uses/.style={-open triangle 45}, node distance = 0.75cm]
	
	\draw[rounded corners, fill=blue25] (-1,0.5) rectangle (1,-7);
	\node[rotate=90]at(0.5,-3.25){datastructures};
	\node[class, text width = 1.8cm, rotate=90,text centered] (ha) at (-0.25,-2.05) {Hybrid automaton};
	\node[class, text width = 1cm, rotate=90, left of= ha, xshift=-1cm,text centered] (point) {Point};
	\node[class, text width = 1.5cm, rotate=90, left of= point, xshift=-1cm,text centered] (hsp) {Halfspace};

	\draw[rounded corners, fill=blue25] (1.5,-5.5) rectangle (12.2,-7.1);
	\draw[rounded corners, fill=blue10] (9.75,-5.6) rectangle (11.6,-7);
	\node[rotate=90] at(11.85,-6.125) {util};
	\draw[rounded corners, fill=blue25] (9,0.5) rectangle (12.2,-3);
	\node[] at(10.75,.25) {algorithms};
	\draw[rounded corners, fill=blue25] (1.5,0) -- (8.5,0) -- (8.5,-3.25) -- (11.85,-3.25) -- (11.85,-5.1) -- (1.5,-5.1) -- cycle;
	\node[rotate=90] at(8.2,-1.5) {representations};
	
	\node[class] (box) at(1.75,-0.5) {Box};
	\node[class, below of = box, node distance=0.6cm] (hpoly) {HPolytope};
	\node[class, below of = hpoly, node distance=0.6cm] (vpoly) {VPolytope};
	\node[class, below of = vpoly, node distance=0.6cm] (pplPoly) {\ppl-Polytope};
	\node[class, below of = pplPoly, node distance=0.6cm] (zono) {Zonotope};
	\node[class, below of = zono, node distance=0.6cm] (sf) {Support function};
	\node[class, below of = sf, node distance=0.6cm,fill=black!10!white] (op) {Orthogonal polyhedra};
	\node[class, below of = op, node distance=0.6cm,fill=black!10!white] (tm) {Taylor model};
	
	\node[class, right of = sf, xshift=3.5cm, yshift=-.5cm, rectangle split, rectangle split parts=2, text centered] (geomObj) 
{\textit{GeometricObject}\nodepart{second}$<$Interface$>$};

	\node[class, text width = 1.5cm,text centered] (converter) at(10,-4.25) {Converter};

	\node[class, text width = 1.1cm] (plotter) at(10,-6.6) {Plotter};
	\node[class, text width = 1.1cm] (logger) at(10,-6) {Logger};
	\node[class, text width = 1cm] (parser) at(10,-0.5) {Parser};
	\node[class, text width = 2cm] (ra) at(9.5,-1.75){Reachability analysis};
	
	\node[class, text width=2cm, text centered] (linOpt) at(4.5,-5.85) {Optimizer};
	\node[class,text centered, below of= linOpt, yshift=-0.1cm,text width=1.1cm, xshift=-2.5cm] (glpk) {\glpk};
	\node[class,text centered, below of= linOpt, xshift=-.75cm, yshift=-0.1cm,, text width = 1.7cm] (smtrat) {\smtrat};
	\node[class,text centered, below of= linOpt, yshift=-0.1cm,, xshift=.875cm, text width = 0.6cm] (z3) {\zthree};
	\node[class,text centered, below of= linOpt, yshift=-0.1cm,, xshift=2.25cm, text width = 1.35cm] (soplex) {\soplex};
	
	\draw[implements] (box) -| ($(geomObj.north) + (-0.25,0)$);
	\draw[implements] (hpoly) -| ($(geomObj.north) + (-0.5,0)$);
	\draw[implements] (vpoly) -| ($(geomObj.north) + (-0.75,0)$);
	\draw[implements] (pplPoly) -| ($(geomObj.north) + (-1,0)$);
	\draw[implements] (zono) -| ($(geomObj.north) + (-1.25,0)$);
	\draw[implements] (sf.east) -- ++ (0.2,0) -- ++(0,-0.25) -- ($(geomObj.west) + (0,0.25)$);
	
	\draw[{open triangle 45-open triangle 45}] (converter.north) |- ($(geomObj.east) + (0,0.375)$);
	
	\draw[uses] (linOpt) -- (smtrat.north);
	\draw[uses] (linOpt) -- (glpk.north);
	\draw[uses] (linOpt) -- (z3.north);
	\draw[uses] (linOpt) -- (soplex.north);
	\draw[uses] (ra) -- (parser);
	
	\draw[<->,double] (1,-2.5) -- (1.5,-2.5);
	\draw[<->,double] (1,-6.25) -- (1.5,-6.25);
	\draw[<->,double] (8.5,-1.5) -- (9,-1.5);
	\draw[<->,double] (12,-3) -- (12,-5.25) -- (11,-5.25) -- (11,-5.6);
	\draw[<->,double] (5.5,-5.1) -- (5.5,-5.5);
	\draw[<->,double] (1,.25) -- (9,.25);
	
\end{tikzpicture}
}
	\caption{\hypro class structure \cite{hypro}.}
\label{fig:structure}
\end{figure}

As mentioned before, there are several state set representations that
can be used in flowpipe-construction-based reachability analysis
algorithms. Hybrid systems reachability analysis tools like, e.g.,
\cora \cite{cora}, 
\flowstar \cite{flowstar}, 
\hycreate \cite{hycreate}, 
\hyreach \cite{hyreachhome}, 
\soapbox \cite{soapbox}, and 
\spaceex \cite{spaceex}
implement different techniques using different geometric
or symbolic state set representations, each of
them having individual strengths and weaknesses. 
For example, \spaceex uses support functions, whereas
\flowstar makes use of Taylor models. 
\newpage
The implementation of state set
representations is tedious and time-consuming, and impedes the (even
prototypical) implementation of new reachability
analysis algorithms.
To offer assistance for rapid implementation, we developed a free and open-source {C++} programming
library \hypro \cite{hypro} (see Figure \ref{fig:structure}), which we
will use in our experiments and which is published at
\url{https://github.com/hypro/hypro}. \hypro contains implementations
for several \emph{state set representations} such as boxes
\cite{moore2009introduction}, convex polytopes
\cite{ziegler1995lectures}, zonotopes \cite{Girard05}, support functions \cite{LeGuernicG10},
orthogonal polyhedra
\cite{bournez1999orthogonal}, and Taylor models \cite{flowstar}, different \emph{operations} on them which are needed
for the implementation of flowpipe-construction-based reachability
analysis algorithms, and \emph{conversions} between the different
representations. \emph{Reduction techniques} can be applied to reduce the representation sizes on the cost of additional over-approximation. 

The implemented representations (with the exceptions of orthogonal polyhedra and Taylor models, depicted grey in Figure \ref{fig:structure}) 
share a \emph{unified interface} to
allow the usage of different representations within a single
algorithm. This property is not only important for extensibility with
new representations but also, e.g., for the implementation of
counterexample-guided abstraction refinement (CEGAR) algorithms: the search
can start with a low-precision but computationally cheap
representation such as boxes, and it can be refined along paths that
are detected to be potentially unsafe by switching to a high-precision but
computationally more expensive representation.

Another important feature of \hypro is that it
is templated in the \emph{number type}, such that it can be
instantiated both with exact as well as with inexact arithmetic.
Linear solver backends such as \glpk\cite{glpk}, \smtrat\cite{smtrat},
\soplex\cite{Wunderling1996}, and \zthree\cite{z3}, which are needed
for the implementation of different operations and conversions, can be
exchanged by the user by her tool of choice.
The library is \emph{thread-safe}, thus parallelisation can be
exploited by the user.  
The efficient usage of the library is further eased
by a \emph{model parsing} module,
a \emph{plotting} engine, and various \emph{debugging} tools.

In this work we illustrate the advantages of the \hypro library by
proposing an algorithm to reduce the computational effort of the search on the cost of
precision loss. Due to space restriction, we do not discuss
refinement steps in this paper, but mention here that using \hypro the
proposed method can be embedded into a CEGAR approach: if a
potentially unsafe path is detected, more precise analysis can be used
to check safety along those paths.
\section{Reachability Analysis based on Variable Set Separation} \label{sec:setSeparation}


\noindent\textbf{Variable set separation and projective representation}\quad
For practically relevant applications, the previously described
modelling approach for PLC-controlled plants by hybrid automata leads to
huge models, even if we exploit the mentioned reduction by
restricted interleaving. The most serious problem is the high
dimensionality: the variable set contains variables modelling the plant
dynamics, the states of sensors and actuators, the input and output
values of the PLC, the local variables used in program executions, and
clocks for PLC cycle synchronisation. The high dimensionality leads to
complex state set representations, causing heavy memory consumption
and computationally expensive applications of state set operations
during the reachability analysis.

To increase scalability and thus to allow the analysis of larger
models, we start with some observations. Firstly,
the variables of the PLC are \emph{discrete}
and thus their values do not change dynamically during time evolution
but only upon taking a discrete transition in the controller part of
the composed hybrid automaton. Furthermore, the
states of actuators and sensors can  be modelled by discrete
variables, as actuator states change discretely (when writing the
output) and the sensor values are relevant only at the beginning of
each cycle (read plant state) as depicted in Figure \ref{fig:plantInterface}. Thus only the physical quantities and the cycle
clocks evolve continuously. 
Finally, computing flowpipes for clocks and other
variables with constant derivatives can usually be done easier than
for dynamics specified by general ODEs.

These observations gave us the idea to divide the variable set $X$
into several disjoint subsets $X=X_1\cup\ldots\cup X_n$ such that
variables in the same subset $X_i$ have some common properties
relevant for reachability analysis.  Once the variables are classified
this way, we could try to modularise the reachability analysis
computation by computing in the sub-spaces defined by the variable
subsets, instead of computing in the global space. However, in order
to compute reachability in the sub-spaces, the variables in different
subsets must be independent in the sense that their evolutions do not
directly influence each other. To be more formal, all predicates
$\varphi\in\Pred{X}$ in the hybrid automaton definition must be
decomposable to a conjunction $\varphi=\varphi_1\wedge\ldots\varphi_n$
of predicates $\varphi_i\in\Pred{X_i}$ over the respective variable
subsets $X_i$, and similarly for jump resets from $\Pred{X\cup X'}$
and flows from $\Pred{X\cup\dot{X}}$. If this condition holds then we
call the subsets $X_1,\ldots,X_n$ themselves as well as variables from
two different subsets \emph{syntactically independent}.

Such a classification of the variable set $X$ into syntactically
independent subsets $X_1,\ldots,X_n$ allows us to represent (global)
state sets $(l,p)\subseteq\Loc\times\mathbb{R}^d$ by their projections
$p\downarrow_{X_i}=p_i\subseteq\mathbb{R}^{|X_i|}$ to the sub-spaces;
we call $(l,p_1,\ldots,p_n)$ the \emph{projective representation} of
$(l,p)$ with respect to the variable separation $X_1,\ldots,X_n$.
Note that the projective representation drops the connection between
the sub-spaces and is therefore \emph{over-approximative}, i.e.,
$p\subseteq p_1\times\ldots\times p_n$ but in general $p\not=
p_1\times\ldots\times p_n$. One exception is the state set
representation by \emph{boxes}: the cross product of the projections
of a box is the box itself, therefore the projective representation of boxes
is exact.

\smallskip
\noindent\textbf{Reachability
  computation based on variable set separation}\quad Given a separation of the variable set $X$ into
syntactically independent subsets $X_1,\ldots,X_n$ and projective
representations based on this separation, we can over-approximate successors of a state set
$(l,p)$ by computing successors of its projective representation
$(l,p_1,\ldots,p_n)$ in each sub-space modularly. As the
computational effort for reachability analysis heavily increases with
the dimension, this modular approach will help to reduce the running
time.  
To explain why we need syntactical independence for sub-space
computations, we first need a more formal description of how successor
sets are computed:\\
\noindent (1) Reachability analysis computes for an initial set $(l,p)$ the
  first flowpipe segment $(l,\Omega_0)$ that over-approximates all states reachable from
  $p$ within a time interval $[0,\delta]$ in $l$ as $\Omega_0 =
  (\textit{conv}(p \cup e^{\delta A}p) \oplus
  \mathcal{B})\cap \Inv(l)$, where the flow in location $l$ is
  $\dot{x}=Ax$, $e^{\delta A}$ is the matrix exponential for $\delta
  A$, $e^{\delta A}p$ are the states reachable from $p$ at time point
  $\delta$, $\textit{conv}(\cdot)$ is the convex hull operator,
  $S_1\oplus S_2=\lbrace a+b\mid a\in S_1\wedge b\in S_2\rbrace$ is the Minkowski sum
  of two sets, the bloating with the box $\mathcal{B}$ accounts for
  the non-linear behaviour between the time points $0$ and $\delta$,
  and $\Inv(l)$ is the invariant for location $l$.\\
\noindent (2) Flowpipe segments $(l,\Omega_i)$ over-approximating the flowpipe within the
  time interval $[i\delta,(i+1)\delta]$ for $i>0$ are computed by
  $\Omega_{i} = e^{\delta A}\Omega_{i-1}\cap\Inv(l)$.\\
\noindent (3) For each jump $e$ with source $l$, guard $g$, reset
  $x'=A'x$ and target location $l'$, each flowpipe segment $\Omega_i$
  is checked for possible successors along $e$ by checking
  $(A'(\Omega_i\cap g))\cap\Inv(l')$ for emptiness. Non-empty
  successors are collected, possibly aggregated, and considered as
  initial state set(s) for location $l'$.

For each decomposition of $X$ into syntactically independent variable sets $X_1,\ldots,X_n$, any
flow $\dot{x}=Ax$ can be decomposed into $\wedge_{i=1}^n\dot{X_i}=A_i X_i$ (where we overload the notation $X_i$ to also denote the sequence of variables in $X_i$). 
Similarly, $\Inv(l)=\wedge_{i=1}^n\Inv(l)_i$ with $\Inv(l)_i\in\Pred{X_i}$.
Furthermore, let $\mathcal{B}_i=\mathcal{B}\downarrow_{X_i}$ be
the 
\begin{wrapfigure}{r}{6.5cm}
	\begin{tikzpicture} [>=latex]
	\draw[thick, ->] (-.5,0) -- (2.5,0);
	\draw[thick, ->] (0,-.5) -- (0,2.5);
	\node[] at (2.25,-.25) {x};
	\node[] at (-.25,2.25) {y};
	
	\draw[thin,dotted] (0.75,-.5) -- (0.75,2.5);
	\draw[thin,dotted] (1.5,-.5) -- (1.5,2.5);
	\draw[thin,dotted] (-.5,0.75) -- (2.5,0.75);
	\draw[thin,dotted] (-.5,1.75) -- (2.5,1.75);

	\foreach \off in {0.75} {
		\draw[blue100] (0+\off,0+\off) -- (.75+\off,.75+\off) -- (.75+\off,1+\off) -- (0+\off,.25+\off) -- cycle;
	}

	\draw[green100, thin] (1.25, -.5) -- (1.25, 2.5);
	\draw[green100, thin] (1, -.5) -- (1, 2.5);
	\draw[green100, fill=green100, opacity=.25] (1, -.5) rectangle (1.25, 2.5);
	\node at (1.125, 0.25) {\small $g$};
	
	\draw[green100, thick, fill=green100, opacity=.5] (1,1) -- (1.25,1.25) -- (1.25, 1.5) -- (1,1.25) -- cycle;
\end{tikzpicture}
	\begin{tikzpicture} [>=latex]
	\draw[thick, ->] (-.5,0) -- (2.5,0);
	\draw[thick, ->] (0,-.5) -- (0,2.5);
	\node[] at (2.25,-.25) {x};
	\node[] at (-.25,2.25) {y};
	
	\draw[thin,dotted] (0.75,-.5) -- (0.75,2.5);
	\draw[thin,dotted] (1.5,-.5) -- (1.5,2.5);
	\draw[thin,dotted] (-.5,0.75) -- (2.5,0.75);
	\draw[thin,dotted] (-.5,1.75) -- (2.5,1.75);

		\draw[blue100] (0.75,-0.05) -- (1.5,-0.05) -- (1.5,0.05) -- (0.75,0.05) -- cycle;
		\draw[green100,fill=green100] (1,-0.05) -- (1.25,-0.05) -- (1.25,0.05) -- (1,0.05) -- cycle;
		\draw[green100,fill=green100] (-0.05,0.75) -- (-0.05,1.75) -- (0.05,1.75) -- (0.05,0.75) -- cycle;
	
	\draw[green100, thin] (1.25, -.5) -- (1.25, 2.5);
	\draw[green100, thin] (1, -.5) -- (1, 2.5);
	\draw[green100, fill=green100, opacity=.25] (1, -.5) rectangle (1.25, 2.5);
	\node at (1.125, 0.25) {\small $g$};
	

\end{tikzpicture}
	\caption{Intersection of a flowpipe segment with an invariant using global (left) and separated (right) variable sets.}
	\label{fig:overapproxError}
\vspace*{2ex}
\end{wrapfigure}
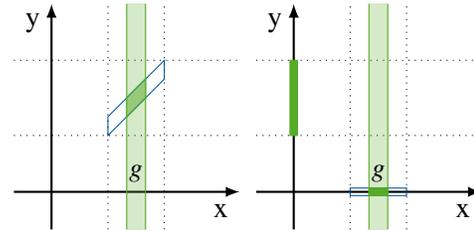
projection of $\mathcal{B}$ to $X_i$. 
Let $(l,p_0,\ldots,p_n)$ be the projective representation of a state set $p$,
\[
\Omega_{0,i}=(\textit{conv}(p_i \cup e^{\delta A_i}p_i) \oplus \mathcal{B}_i)\cap \Inv(l)_i 
\]
and for each $j>0$ 
\[
  \Omega_{j,i}=(e^{\delta A_i}\Omega_{j-1,i}%
)\cap \Inv(l)_i\, .
\]
Then $\Omega_j\subseteq \Omega_{j,1}\times\ldots\times\Omega_{j,n}$.

The computations in the sub-spaces are precise as long as the initial
set $p$ is a box and the flowpipe resides inside the invariant, i.e.,
if $p$ is a box and $\Omega_{m,i}\subseteq \Inv(l)_i$ for all $1\leq
m\leq j$ and all $1\leq i \leq n$ then
$\Omega_j\downarrow_{i}=\Omega_{j,i}$.

However, syntactical independence does not imply semantical
independence, as the different dimensions are usually still implicitly
connected by the passage of time. If one of the projections runs out of a non-trivial
invariant then the intersection with the (projection of the) invariant in a sub-space does not
necessarily affect the computations in the other sub-spaces, thus the
result might become over-approximative (see Figure
\ref{fig:overapproxError}). To increase precision, we can at least
incorporate that if the projection of a flowpipe segment gets empty in
one of the sub-spaces then the whole flowpipe segment gets empty:
instead of $\Omega_{j,i}$ we use $\Omega_{j,i}'$ that is
$\Omega_{j,i}$ if none of $\Omega_{j,k}$, $k=1,\ldots,n$ is empty and
the empty set otherwise.

For successors along jumps, the reachability computations in the
sub-spaces work similarly. Also here,
additional over-approximation might be introduced by intersections
with guards and invariants in target locations, which we try to reduce
by the above-described emptiness check.

As we use modular computations in sub-spaces, on the one hand our
method speeds up reachability computations, but on the other hand it introduces
additional over-approxima\-tions.
Therefore, in our experiments we will thoroughly analyse the effect of
our approach both to the running time as well as to the
over-approximation error.  Besides
the reduced computational effort, our method has further advantages.
For example, state sets in the sub-spaces can be represented
independently of each other, using different state set
representations. We observed that the discrete variables
can often be represented by boxes without serious over-approximation,
whereas the plant dynamics requires a more precise representation,
e.g. by support functions. Furthermore, for sub-spaces defined by
clocks or by variables with constant derivatives one could use
different, computationally less expensive techniques for computing
flowpipes.

\begin{algorithm}
\scalebox{0.9}{
\begin{minipage}{1.04\linewidth}
  \KwIn{Hybrid system model $H=(\Loc,\Var,\Flow,\Inv,\Edge,\Init)$, time step $\delta\in\mathbb{Q}_{\geq 0}$, global time 
horizon $T\in\mathbb{Q}_{\geq 0}$, jump depth $D\in\mathbb{N}_{\geq 0}$, aggregation flag $\textit{aggregation}\in\{0,1\}$.}
  \KwOut{An over-approximation $\{(l,x)|(l,p_{\disc},p_{\clock},p_{\rest},[t_1,t_2])\in R\wedge x\in p_{\disc}\times p_{\clock}\times p_{\rest}\}$ of the states reachable within the given bounds.}
  $\dVar := $ maximal set of variables from $\Var$ with derivatives 
$0$ that are syntactically independent from $\Var\setminus\dVar$\;\label{alg:varSepBegin}
  $\cVar := $ maximal set of variables from $\Var$ with derivatives $1$ that are syntactically independent from $\Var\setminus\cVar$\;
  $\rVar := \Var\setminus(\dVar\cup\cVar)$\;\label{alg:varSepEnd}
  choose a representation type for each of $\dVar$, $\cVar$, and $\rVar$\;\label{alg:repr}
  bring each predicate $\varphi$ in $H$ to an equivalent form $\varphi_{\disc}\wedge \varphi_{\clock}\wedge \varphi_{\rest}$, 
where each $\varphi_{i}$, $i\in\{\disc,\clock,\rest\}$, is a predicate from $\Pred{\Var_{i}}$ resp. 
$\Pred{\Var_{i}\cup\Var'_{i}}$ (jump resets) resp. $\Pred{\Var_{i}\cup\dot{\Var}_{i}}$ (flows)\;
$P:=\emptyset$; $R := \emptyset$\; 
  \ForEach{location $l\in\Loc$}{\label{alg:initSetCheckBegin}
    let $p_i := \Init(l)_i\cap \Inv(l)_i$ for each $i\in\{\disc,\clock,\rest\}$\;
    \If{$p_{\disc}\not=\emptyset\wedge p_{\clock}\not=\emptyset\wedge p_{\rest}\not=\emptyset$}{add $(l,p_{\disc},p_{\clock},p_{\rest},[0,0])$ to $P$}
  }\label{alg:initSetCheckEnd}
  \While{$P\not=\emptyset$}{\label{alg:loopBegin}
    choose $p=(l,p_{\disc},p_{\clock},p_{\rest},[t_1,t_2]) \in P$ and remove $p$ from $P$\;\label{alg:selectState}
    $E := \{(e,p_{\disc}^e) \ |\  e=(l,g,r,l')\in\Edge\wedge p_{\disc}^e=\textit{jump}(p_{\disc}, g_{\disc}, 
r_{\disc}, \Inv(l')_{\disc})\not=\emptyset\}$\; \label{alg:discreteJumpCheck}
    $\isFirst := \textit{true}$;
    \lForEach{$(e,p_{\disc}^e)\in E$}{
      $P^e:=\emptyset$
    }\; \label{alg:first}
    \While{$\textit{true}$}{\label{alg:flowLoopBegin}
    \tcp{Compute flowpipe, considering also the invariant of the location}
      \If{$t_1 < T$\label{alg:computeTime}}{
        $p_{\clock} := \textit{flow}(p_{\clock},\Flow(l)_{\clock},\Inv(l)_{\clock},\delta,\isFirst)$;
        \lIf{$p_{\clock}=\emptyset$}{break}\;
        $p_{\rest} := \textit{flow}(p_{\rest},\Flow(l)_{\rest},\Inv(l)_{\rest},\delta,\isFirst)$;
        \lIf{$p_{\rest}=\emptyset$}{break}\;
        $t_2 := t_2 + \delta$; \lIf{$\neg\isFirst$}{$t_1 := t_1 + \delta$}\;
      }\label{alg:computeTimeEnd}
      $R := R\cup \{(l,p_{\disc},p_{\clock},p_{\rest},[t_1,t_2])\}$\;
      \tcp{Compute jump successors}
      \ForEach{$((l,g,r,l'),p_{\disc}^e)\in E$\label{alg:computeJmp}}{
        $p_{\clock}^e:=\textit{jump}(p_{\clock}, g_{\clock}, r_{\clock}, \Inv(l')_{\clock})$;
        \lIf{$p_{\clock}^e=\emptyset$}{continue}\;
        $p_{\rest}^e:=\textit{jump}(p_{\rest}, g_{\rest}, r_{\rest}, \Inv(l')_{\rest})$;
        \lIf{$p_{\clock}^e=\emptyset$}{continue}\;
        add $(l',p_{\disc}^e,p_{\clock}^e,p_{\rest}^e, [t_1, t_2])$ to $P^e$\; 
      }\label{alg:computeJmpEnd}
      $\isFirst := \textit{false}$; \lIf{$t_1\geq T$}{break}\;
    }\label{alg:flowLoopEnd}
    \ForEach{$e\in E$}{
      \If{$P^e\not=\emptyset$}{
        \lIf{\textit{aggregation}}{$P := P\cup\{\textit{aggregate}(P^e)\}$;} \label{alg:aggreg}
        \lElse{$P := P\cup P^e$}
      }
    }
  }\label{alg:loopEnd}
  \KwRet{$R$}
\end{minipage}
}
  \caption{Reachability analysis algorithm based on variable set separation for hybrid automata.}
\label{alg:reachability}
\end{algorithm}

\smallskip
\noindent\textbf{The algorithm}\quad
Our reachability analysis algorithm based on variable set separation is
presented in Algorithm \ref{alg:reachability}. 
The input is a hybrid automaton $H$, a parameter $\delta$ that
specifies the length of a time step in the flowpipe construction, a global
time horizon $T$ and a jump depth $D$ that specify upper bounds on the
total time duration respectively on the total number of jumps in the
considered execution paths, and an aggregation flag that specifies
whether aggregation should be applied to the successors of segments of
a flowpipe along a jump. The algorithm outputs a set of flowpipe segments $R$, where the union of the segments in $R$ is an 
over-approximation
of the set of states that are reachable in $H$ within the given time
and jump bounds.

In a preprocessing step, we separate the variable set into three
syntactically independent subsets of discrete variables, clocks, and
the rest (lines \ref{alg:varSepBegin}-\ref{alg:varSepEnd}). We have
chosen this variable set separation as it seems to
be practically helpful in our experiments, but other separation
criteria could be defined, too. A state set representation can be
chosen for each sub-space independently (line \ref{alg:repr}); for
readability, in the algorithm we do not distinguish between state sets
and their representations syntactically.

The invariant conditions for the initial states are checked for each
initial set and each sub-space independently in the lines
\ref{alg:initSetCheckBegin}-\ref{alg:initSetCheckEnd} and if the
intersections of the initial sets with the invariants are non-empty in
all sub-spaces then they are added to a set $P$ of state sets, whose
successors need to be determined. Additionally to the location and the
projective representation of state sets, we attach to the state sets
the time interval within which the represented states can be reached.

As long as $P$ is not empty, we choose a state set $p\in P$.  Before
computing its flowpipe, we determine the set of jumps with $p$'s
location as source. As the values of discrete transitions do not
change during time evolution, we can skip those jumps whose guard is
violated by the initial values of the discrete variables, and store
the remaining ones in the set $E$ (line \ref{alg:discreteJumpCheck}).
The sets $P^e$ will be used to collect non-empty successors from
different flowpipe segments along the jumps $e\in E$.

Next we compute the segments of $p$'s flowpipe as explained previously
(lines \ref{alg:computeTime}-\ref{alg:computeTimeEnd}).
Instead of one single flowpipe of dimension $d$, our algorithm
computes lower-dimensional flowpipes in the sub-spaces using a common
time step size and a global time horizon (line \ref{alg:computeTime}) to be
able to connect the flowpipe segments computed in the sub-spaces. The variable $\isFirst$, initialised in
line \ref{alg:first}, is used to remember
whether the flowpipe segment to be computed next is the first one (as
the first one needs special handling).

Once the flowpipe segments are computed, their jump successors are
determined and collected in the sets $P^e$ for each outgoing jump $e\in
E$ (lines \ref{alg:computeJmp}-\ref{alg:computeJmpEnd}), if aggregation is activated then they are aggregated (per jump, line \ref{alg:aggreg}), and finally added to $P$ for further iterative successor computation.
Note that the flowpipe as well as the jump successor computations in the sub-spaces
are time-synchronised: if a successor set gets empty in one of the sub-spaces then
the computation is stopped.

\section{Experimental Results} \label{sec:experiments}

All computations were carried out on an Intel Core i7 CPU with 8 cores and 16 GB RAM. We used the time step size $\delta=0.01$ and unlimited 
jump depth. To be able to express a global time horizon, each model is equipped with a global clock.

\begin{table}[t]
	\caption{Model sizes of the benchmarks.}
	\centering
\vspace*{-2ex}
\scalebox{0.85}{
	\begin{tabular}{lcrrrrrr}
		\toprule
		Benchmark							& Type				& \multicolumn{3}{c}{\#variables}	& \multicolumn{2}{c}{\#modes}	& \#jumps\\
 & & disc. & clocks & rest & controller	& plant	& \\
		\midrule
		\multirow{4}{*}{Leaking tank}		& original				& 0		& 0 & 12											& 8			
	& 3										& 34\\
											& timed					& 0		& 2& 10										& 8				& 3										& 34\\
											& discrete						& 9		& 0	& 3									& 8				
& 3										& 34\\
											& timed \& discrete		& 9		& 2	& 1										& 8				& 3	
									& 34\\
		\midrule				
		\multirow{4}{*}{Two tanks}			& original				& 0		& 0& 22											& 20			
& 14									& 296\\
											& timed					& 0		& 3& 19										& 20			& 14									& 296\\
											& discrete						& 17	& 0& 5										& 20		
	& 14									& 296\\
											& timed \& discrete			& 17	& 3& 2										& 20			
& 14									& 296\\
		\midrule
		\multirow{4}{*}{Thermostat}			& original				& 0		& 0	& 8 										& 6				
& 2										& 18\\
											& timed						& 0		& 2& 6										& 6				& 2										& 18\\
											& discrete						& 5		& 0& 3										& 6			
	& 2										& 18\\
											& timed \& discrete			& 5		& 2	& 1									& 6				& 2	
									& 18\\
		\bottomrule
	\end{tabular}
}
	\label{tab:bmStatistics}
\end{table}

\smallskip\noindent\textbf{Benchmarks}\quad
For our experiments we used three well-known benchmarks, which we
slightly modified by
adding model components for PLC controllers. Besides increasing the
number of modes, these extensions add variables with discrete
behaviour (i.e. with zero derivatives) to
model the actuators and sensors of the plant and the
input, output, and local variables of the controller. Furthermore, one clock variable is added
for each introduced PLC controller to model the cycle time, and one discrete variable to store
the controller mode. In our experiments we compare the analysis of the
benchmarks without variable separation (``original'') with variable-set-separation-based analysis
separating only clocks (``timed''), only discrete
variables (``discrete''), and both (``timed \& discrete''). The sizes of the models are shown
in Table~\ref{tab:bmStatistics}. The modified versions of the
benchmarks are accessible as part of our benchmark collection \cite{benchmarksite}.
A binary of our implementation can be found at
\cite{hyprosite}.

\smallskip
\noindent\emph{Leaking tank}\quad
This benchmark models a water tank which leaks, i.e., it has a
constant outflow. The tank can be refilled from an unlimited external
resource with a constant inflow that is larger than the outflow. The
PLC controller triggers refilling (by switching a pump on) if a sensor
indicates a low water level ($h \leq 6$). If the water level is high ($h
\geq 12$) the controller stops refilling (switches the pump off).
Adding the controller introduces two controller input variables for
low and high water levels, variables for the actuator (pump) state in
the plant and the controller, and a variable to store the controller
mode.  Furthermore, a new clock is added to model the PLC cycle time.
Besides the controller we also model a user which can manually switch
the pump on and off as far as the water level allows it. In our implementation, the user constantly
toggles between the pump states on and off. We analyse the system
behaviour over a global time horizon of $40$ seconds using a PLC cycle
time of $2$ seconds.
 
\begin{table}[t]
	\caption{Benchmark results for different separation set-ups. Running times are in 
	seconds, time-out (TO) was 20 minutes, in brackets we list the number of flowpipes computed.}
	\centering
\vspace*{-2ex}
\scalebox{0.85}{
	\begin{tabular}{lccrrrrr}
								\toprule
								& 	   &        	& \multicolumn{4}{c}{\hypro} 					& \spaceex \\
		Benchmark 				& Rep. & Agg		& original 		& timed  		& discrete  & timed \& discrete	& original\\
								\midrule
\multirow{4}{*}{Leaking tank}	& box  & agg    	& 2.70 (662) 	& 2.08 (662) 	& 1.06 (662) & 1.13 (662)  	& 3.67 (200)\\
								& box  & none 	& 2.62 (662) 	& 2.09 (662) 	& 1.06 (662) & 1.13 (662)  	& 3.82 (200)\\
								& sf   & agg    	& TO (18) & TO (28) & 161.12 (662) & 37.03 (662)	&448.3 (425)\\
								& sf   & none 	& TO (583) & 1044.97 (662) & 19.49 (662) & 5.84 (662)	&444.82 (425)\\
								\midrule
\multirow{4}{*}{Two tanks}		& box  & agg    	& 4.39 (470) 	& 2.60 (470) 	& 0.97 (470) & 1.15 (470)  	& 5.49 (195)\\
								& box  & none 	& 4.46 (470) 	& 2.68 (470) 	& 1.02 (470) & 1.16 (470)   & 5.53 (195)\\
								& sf   & agg    	& TO (4) & TO (4) & 900.11 (470) & 329.80 (470) &TO (171)\\
								& sf   & none 	& TO (54) & TO (64) & 35.04 (470) & 14.64 (470) &TO (172)\\
									\midrule
\multirow{4}{*}{Thermostat}		& box  & agg    	& 0.07 (95) 	& 0.09 (95) 	& 0.06 (95) & 0.06 (95) 	& 0.57 (95)\\
								& box  & none 	& 0.11 (95) 	& 0.09 (95) 	& 0.06 (95) & 0.06 (95) 	& 0.57 (95)\\
								& sf   & agg    	& 35.87 (95) 	& 22.69 (95) 	& 1.17 (95) & 0.29 (95)		& 9.89 (84)\\
								& sf   & none 	& 30.41 (95) 	& 20.19 (95) 	& 1.18 (95) & 0.30 (95) 	& 9.91 (84)
\\
								\bottomrule
	\end{tabular}
}
	\label{tab:results}
\end{table}

\smallskip
\noindent\emph{Two tanks}\quad
This benchmark models the water levels of two water tanks in a closed
system. Each tank has a constant inflow and a constant outflow. The
tanks are connected via pipes, such that the amount of water outflow
of the first tank is equal to the inflow of the second tank and vice
versa. One pump per pipe allows to enable/disable the water flow. We
add a controller to the two tank system that controls the pumps. A
pump is switched off if the water level of the source tank is low ($h
\leq 8$) or if the water level of the target tank is high ($h \geq
32$). Each time a pump is switched off by the controller, the other
pump is switched on to balance the water levels in the tanks.  The
introduction of the controller adds variables to model sensing low and high water levels
of both tanks and variables to model
the actuator (pump) states in the plant and the
controller. Moreover, we add a variable to store the controller mode
and a new clock to model the PLC cycle time.  Again, we model a user
which switches the pumps manually on or off as far as the water levels allow it. We implemented a user
that toggles the state of each pump in each PLC cycle. The global time
horizon and a PLC cycle time were set to $20$ seconds respectively $1$ second.

\smallskip
\noindent\emph{Thermostat}\quad
In this benchmark a heater with a thermostat controller is modelled.
Initially, the temperature is $\temperature = 20^\circ C$ and the
heater is on.  The controller keeps the temperature $\temperature$
between $16^\circ C$ and $24^\circ C$.  The heater is switched off if
the temperature rises above $23^\circ C$ and it is switched on at a
temperature below $18^\circ C$. Adding
a controller to the model introduces new variables for the low and high temperature
sensors in the controller, a variable for
the actuator (heater) state in the plant and the controller, and a
variable to store the controller mode. Additionally, we introduce a
new clock for the cycle time of the PLC. The global time horizon is
$10$ seconds and the PLC cycle time is $0.5$ seconds.



\begin{figure}
\begin{center}
	\includegraphics[width=0.49\textwidth]{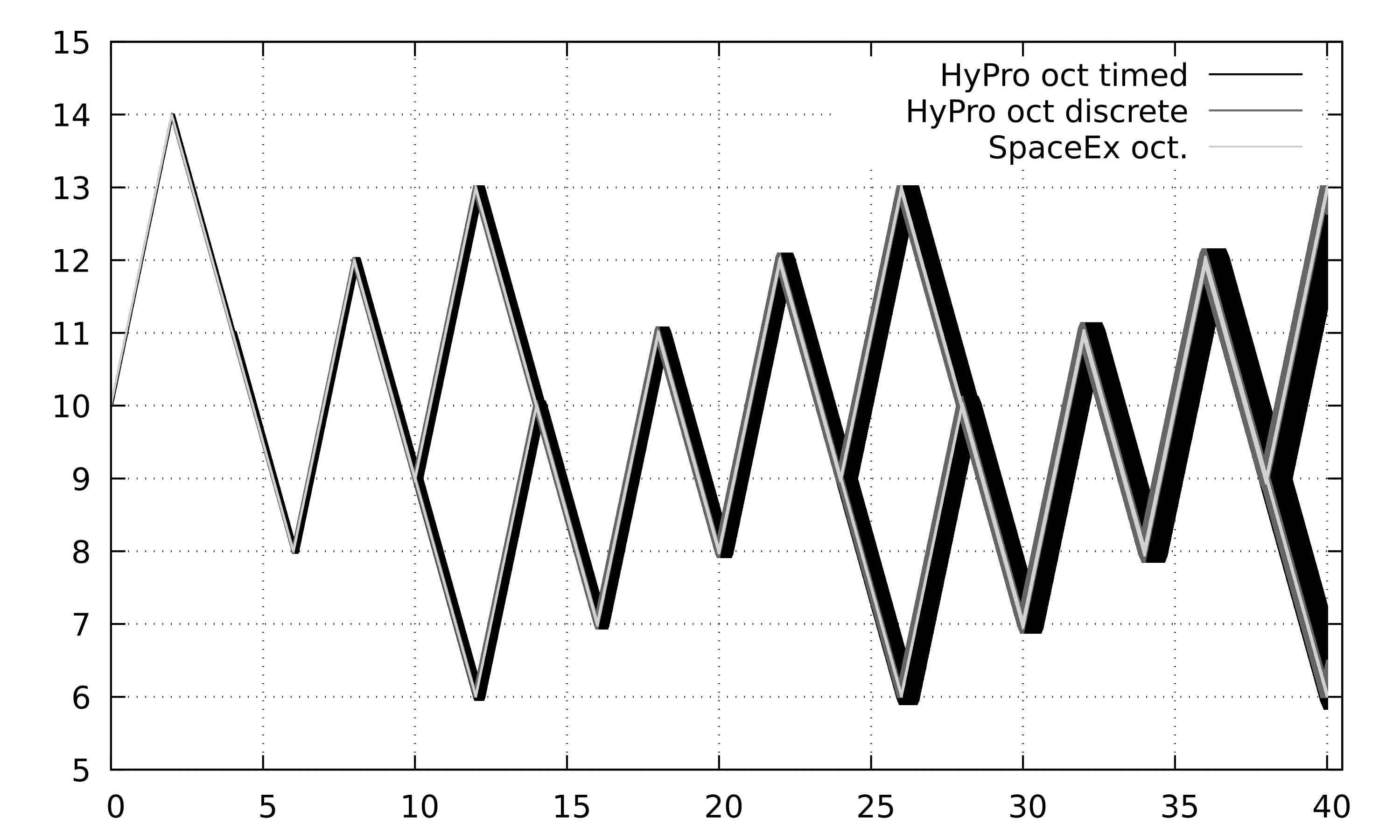}
	\includegraphics[width=0.49\textwidth]{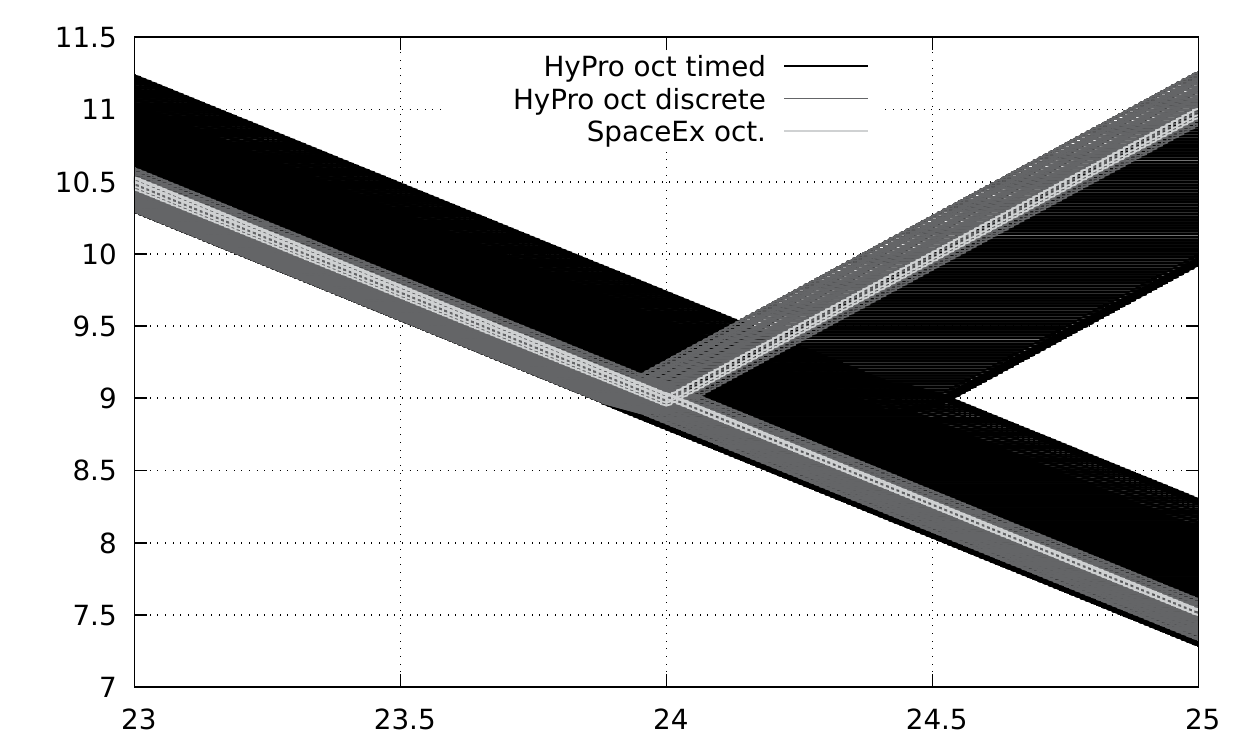}
\end{center}
\vspace*{-2ex}
	\caption{\spaceex and \hypro results on the leaking tank
          benchmark with support function representation (using a regular octagonal (oct) template for evaluation), when \hypro separates 
either only clocks or  only
          discrete variables.}
        \label{fig:leakingTankBig}
\end{figure}

\smallskip
\noindent\textbf{Results}\quad
We implemented our algorithm using the \hypro library and evaluated it
on the above benchmarks. Table \ref{tab:results} shows results from our tool and \spaceex version $0.9.8$f. In our tool we 
used
boxes and support functions (evaluated in 8 directions) to represent state sets,
whereas in \spaceex we used support functions with
$4$ and $8$ directions, as \spaceex does not support explicit box
representations.

The \hypro and \spaceex results are not fully comparable because
\spaceex implements a fixed-point detection algorithm but \hypro does
not. The leaking tank benchmark as well as the two tank benchmark both
cause branching in the execution paths which are merged later (see
Figure \ref{fig:leakingTankBig}). Our implementation does not
recognise the merging of these paths and fully computes each branch
independently.  Thus \hypro needs to compute a higher number of
flowpipes (given in brackets behind the running times in Table
\ref{tab:results}) than \spaceex.  Another difference is that in
\hypro we varied the state set representation for the continuous sets
between boxes and support functions (similarly to \spaceex) but used
boxes for the discrete and the clock variable sets in all settings.

\begin{wrapfigure}{r}{8cm}
\vspace*{-2ex}
	\includegraphics[width=\linewidth]{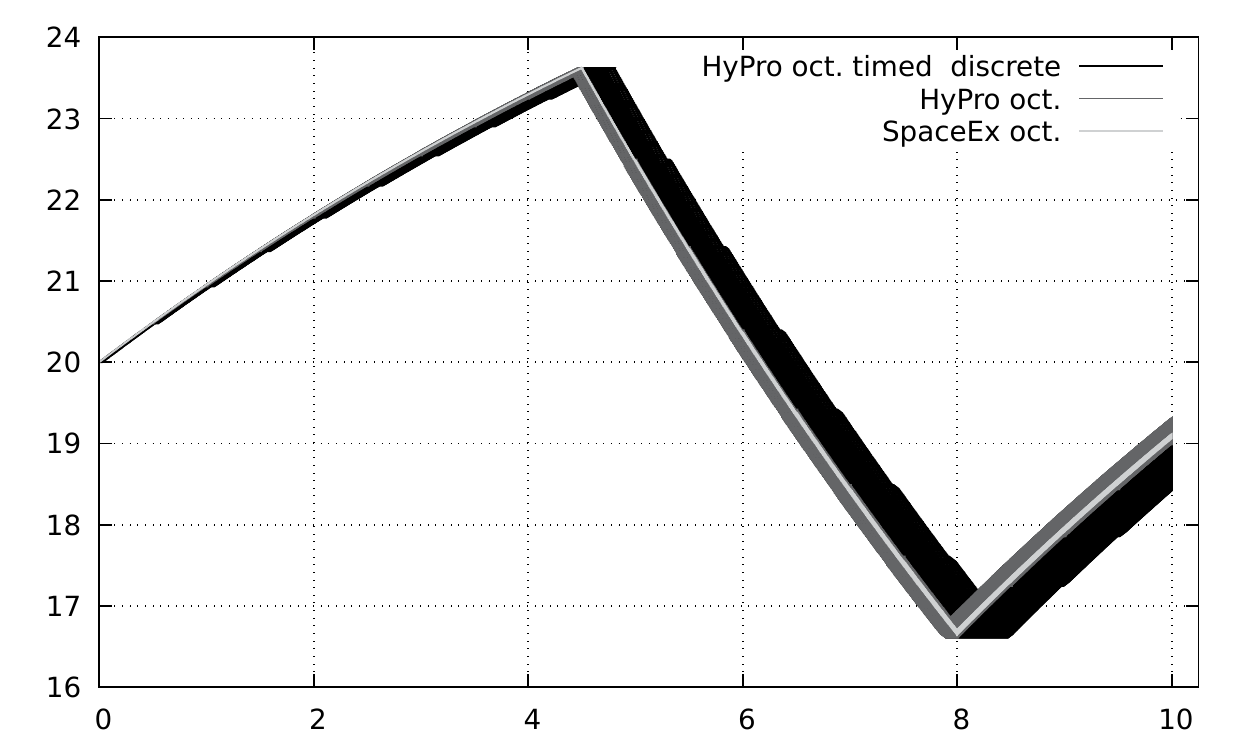}
	\caption{\spaceex and \hypro results on the thermostat
          benchmark with support function representation (using an octagonal (oct) template for evaluation), when
          \hypro separates clocks as well as discrete variables.}
        \label{fig:thermostatExtract}
\vspace*{-2ex}
\end{wrapfigure}
Using variable separation clearly improves the running times, due
to computations in lower-dimensional sub-spaces.  However,
we can also observe on the Figures \ref{fig:leakingTankBig} and
\ref{fig:thermostatExtract}, which show plots for the detected
reachable regions for the leaking tank and the thermostat, that separating the
clock variables (which measure the cycle time and the global time)
introduces a slight over-approximation.

The influence of the discrete variable
separation is in general larger than the influence of a clock
separation, probably because in our benchmarks the discrete variables outnumber the
clocks. Nonetheless a separation of clocks already
shows a speed-up of about $30\%$.  As mentioned before, we used boxes
as a state set representation for the set of discrete variables, which
does not introduce any further over-approximation error, as the discrete
variables themselves are all syntactically independent. We can observe
that using boxes as a state set representation, our implementation
outperforms \spaceex (even when a lot more flowpipes are computed),
which is expected, as boxes in general require less computational
effort than support functions (evaluated in $4$ directions) in reachability analysis. 

In \hypro, aggregation causes longer running times because
in the current implementation
aggregation is realised by a conversion of the single sets
(which are to be aggregated) to polytopes, which is computationally
expensive, especially in higher dimensions.

\section{Conclusion} \label{sec:conclusion}

In this paper we presented an approach to reduce the computational
effort in the reachability analysis of hybrid systems for certain
applications. Our
experimental results indicate that even state-of-the art reachability
analysis tools struggle to analyse high-dimensional models with
relatively simple dynamics, which are common in the application area of controlled plants.

In general controlled plants are composed of many single components such as the set of controllers or the physical quantities of the plant. 
A naive approach models each of these components and the full model is the result of a parallel composition of the 
single components. Even the relatively simple examples used in Section \ref{sec:experiments} yield large models which put 
state-of-the-art reachability analysis tools to their limits. To increase scalability, domain-specific knowledge helps to create more 
sophisticated and smaller models. For example knowing that PLC computation as well as the plant's 
behaviour do not interfere during a PLC cycle already allows to prohibit arbitrary switching between the controller and the plant, which reduces the  model 
complexity.

In contrast to common benchmarks for hybrid systems, our plant models exhibit a large number of discrete variables accounting for the 
controller's behaviour. Currently available tools do not distinguish between the different dynamics of variables, thus discrete variables 
usually are treated as continuous variables and unnecessarily increase the dimension of the state space.

Our approach allows to split variable sets according to
their dynamics, which has a positive effect on the running times, as
reachability analysis algorithms can be protected from working in
high-dimensional spaces. We can observe that the distribution of variables to the different sets has a high influence on the computation time. We 
expect that splitting the set of continuous variables (if applicable) into multiple, independent sets with fewer variables each will result 
in the best results regarding computation time. Furthermore in applications with several PLC controllers, each control program operates 
independently, which allows to build separate variable sets for each controller. Depending on the dimension of the individual sets and the 
associated dynamics for the contained variables, utilizing individual state set representations can be 
beneficial. As state sets for independent discrete variables are always hyper-rectangles, using boxes instead of other, computationally more expensive
state set representations has shown great improvements in terms of runtime. For higher dimensional state sets support 
functions can be expected to perform better than other state set representations.

In our application scenario, syntactically independent variable sets are directly given. In general, this is not necessarily the case for 
hybrid system models. Transforming the state space can help to identify independent variable sets and allows to apply the presented 
approach to systems where the independent variable sets are not obvious.

As to future work, we will improve our implementation by adding fixed-point detection and a more sophisticated implementation for state set aggregation. 
Second, we will embed the presented approach into a CEGAR framework to refine potentially unsafe paths. Finally, we also work on 
parallelisation approaches for flowpipe computations.

\bibliographystyle{eptcs}
\bibliography{references}
\end{document}